\documentclass[11pt,twoside]{article}
\usepackage[pdftex]{graphicx}
\usepackage{amsmath}
\usepackage{amssymb}
\usepackage{cite}

\begin{document}
\newcommand{\pst}{\hspace*{1.5em}}


\newcommand{\be}{\begin{equation}}
\newcommand{\ee}{\end{equation}}
\newcommand{\bm}{\boldmath}
\newcommand{\ds}{\displaystyle}
\newcommand{\bea}{\begin{eqnarray}}
\newcommand{\eea}{\end{eqnarray}}
\newcommand{\ba}{\begin{array}}
\newcommand{\ea}{\end{array}}
\newcommand{\arcsinh}{\mathop{\rm arcsinh}\nolimits}
\newcommand{\arctanh}{\mathop{\rm arctanh}\nolimits}
\newcommand{\bc}{\begin{center}}
\newcommand{\ec}{\end{center}}

\thispagestyle{plain}

\label{sh}


\begin{center} {\Large \bf

"NOBODY UNDERSTANDS QUANTUM MECHANICS." WHY?
 } \end{center}

\bigskip

\bigskip

\begin{center} {\bf
Nicolay V. Lunin

}\end{center}

\medskip

\begin{center}
{\it
Volga State Academy of Water Transport\\ 
Nizhny Novgorod, Russia\\

e-mail:~~~n.v.lunin@sandy.ru
}
\end{center}

\begin{abstract}\noindent

The well known and oft-quoted Feynman's  expression, entered the title, leading at a loss and even being objectionable, has not yet a clear explanation.
 
The hidden parameters problem in quantum mechanics is considered here on the base of group-theoretic approach which includes the complete set of observables indispensably. The last ones are the bilinear Hermitian forms constructed from the Schroedinger equation solutions and its first derivatives, they satisfy the algebraic completeness condition.

These Hermitian forms, obtained for the simplest standard problem of particle transmission above potential step, had been compared with the Hermitian forms which are usually considered in this problem, and an additional ones, which may be obtained within the framework of an ordinary schemes of quantum mechanics.

It is shown that the generally recognized schemes of the problem solution lead to violation of some conservation laws on the step directly. 

On the contrary, the group-theoretic approach leads to fulfillment of all necessary conservation laws everywhere at the same time. 

It is also shown that the complete set of observables leads a probabilistic interpretation in quantum mechanics to be excessive.

\end{abstract}

\medskip


\section{Introduction}
\pst

      The Feynman assertion in the title of present paper, expressed half a century ago
\cite{1}, and evoking some perplexity, may be considered as the direct consequence of the well known question formulated more earlier. The last one, "Can quantum-mechanical description of physical reality be complete?", entered the title of \cite{2}, is remained to be unsolved and exciting up to now \cite{3}. Together with other Feynman's remarks that the uniqueness of probabilistic wave function interpretation in quantum mechanics is not yet proved, and even it is a problem "to show that the probability interpretation of  $|\psi|^2$ is the {\it only} consistent interpretation of this quantity" \cite{4}, they give us sufficient reasons to consider this questions once more.

    Close connection of wave function with probability density engages an exclusive role in quantum mechanics. It is explained mainly by the fact that
"the details of this have been analyzed only on the assumption that 
 $|\psi|^2$ is a probability, and the consistency of this assumption has been shown. It would be an interesting problem to show that {\it no other} consistent interpretation can be made" \cite{4}. Similar point of view is contained also in such significant books on quantum mechanics foundations as \cite{5} and \cite{6}.

  Extremely high status of probability density $\rho = \psi \psi^*$ in quantum mechanics allows one to wait that its consequences have to lead to whole clarity, understanding and common point of view on probability and maybe on hidden parameters  in some kind of extremely simplest problem of quantum mechanics. If sufficient level of clarity can not be achieved in the simplest problem of quantum mechanics then what may one say in another cases?

   To solve such kind problems it seems to be well founded to accept some undeniable facts as an initial point of consideration. Here we shall accept fulfillment of the conservation laws as an experimental fact, the group theory and the Noether theorems as a mathematical tool for its realization, and the Schroedinger equation as far as this one describes many quantum objects and phenomena successfully.

    Besides, it is necessary to find the extremely simple quantum problem which is the simplest one from physical point of view, on the one hand, and also the simplest one from calculating point of view, on the other hand.

     It seems that such requirements are satisfied in the case of the 
simplest, comprehensively  investigated and maybe standard problem, simpler of which it does not exist in quantum mechanics, that is the problem on passage of particle described by the unidimensional stationary Schroedinger equation above potential step. This problem is considered here from two points of view, which may be named universally recognized approach as the first one, and the group-theoretic approach as the second one, with the purpose of their subsequent comparison.

    The group theory significance for the problem mentioned above is connected beforehand with the Noether theorems. Last ones establish one-to-one correspondence of solution's transformations groups for equation describing physical phenomenon with necessary conservation laws \cite{7}.
It is also important that the group theory describes symmetries of scalar and vector variables which may be connected with energy, linear and angular momentums measured in experiment. These circumstances allow one to be sure that the physical theory constructed as consecutive group-theoretic scheme will satisfy all necessary requirements to observables.

  From the physical point of view, the theoretical results have to be compared with experimental measurements, therefore it is necessary to construct  mathematical variables which are measurable in principle, on the one hand, and which may be associated with variables observed in experiment, on the other hand. The Hermitian forms constructed on the base of the Schroedinger equation solutions or its spinor representation are usually used as observables. For example,  a "probability density" 
$\psi \psi^*$ mentioned above is such variable.

    Ascertainment of transformation properties of observables, their connections among themselves, conservation laws fulfillment and comparison of the theoretical conclusions with experimental data in general are {\it unthinkable} without definition of the Hermitian forms complete system. In particular it means that the Hermitian forms have to satisfy some algebraic completeness condition.

      The group-theoretic approach is based on definition of propagators group-theoretic belonging, ascertainment of topological properties of propagators transformations space for the Schroedinger equation spinor representation and definition of the Hermitian forms complete system constructed either on the base of wave function and its derivative or on the base of spinor components \cite{8,9}.

       The second part of paper is devoted to description of the general peculiarities of group-theoretic approach essential for the problems to be considered.

       A detailed calculations of the problem on potential step in accordance with group-theoretic approach are presented in the third part of paper.

       The forth part contains presentation of the same problem solution as it had been considered in \cite{6}. This solution of the problem may be considered as the general accepted one.

       The results obtained in two previous parts of the paper are compared in the fifth part  to select and emphasize differences between two approaches
with respect to some essential problems of quantum mechanics.

       The last part, Conclusion, contains some consequences and inferences which may be carried out from comparison of results of two different approaches applied to the simplest problem of quantum mechanics. It would be expected that some part of differences may be occurred to be responsible with respect to the question in the title.

\section{Peculiarities of the group-theoretic approach}
\pst

    Let us set forth some known facts on a group-theoretic approach essential for the problem to be considered. 

    Aiming the purpose to approach a group-theoretic description, let us at first go over, accordingly to
\cite{10,11,12},
 from the unidimensional stationary Schroedinger equation for complex functions 

\begin{equation}\label{1}
\psi^{''}(z) + k^2(z) \psi (z) = 0
\end{equation}
to pair of first order equations for functions $\Phi_{\pm}$ connected with wave function and its derivative by means of following equalities
 
\begin{equation}\label{2}
\Phi_{\pm}(z) = \frac{k^{1/2}(z)}{\sqrt{2}} [\psi (z){\pm} \frac{1}{ik(z)} \psi^{'}(z)].
\end{equation}

    These equations may be written in matrix form as

\begin{equation}\label{3}
\Phi^{'}(z) = [ik(z)\sigma_3 + \frac{k^{'}(z)}{2k(z)}\sigma_1]\Phi(z)
\end{equation}
for column

\begin{equation}\label{4}
\Phi (z) =
\Big|\Big| \begin{array}{c} \Phi_+(z)\\{\Phi_-(z)}
\end{array}\Big|\Big| =
\Big|\Big|\begin{array}{c}
a e^{i\alpha}\\
b e^{i\beta}\\
\end{array}\Big|\Big|
\end{equation}
with arbitrary conditions at the initial point $z_0$

\begin{equation}\label{5}
\Phi_+(z_0) = a_0 e^{i\alpha _0}, \quad
\Phi_-(z_0) = b_0 e^{i\beta _0}. 
\end{equation}{}

   The equation (3) is a spinor representation of the Schroedinger equation,  
it allows one to use matrix  representations of groups to investigate transformation properties of propagators for solutions and conservation laws accordingly to the Noether theorems 
\cite{7}.

    To compare conclusions of any scheme of quantum theory with experimental data one needs to construct a real measurable variables based on complex solutions of the Schroedinger equation. A combinatorial analysis 
leads to conclusion that only four Hermitian forms may be constructed on basis of wave function together with its derivative, or in terms of two spinor components, coupled with their complex conjugate ones, of course. Here
we accept them in the following form

\begin{equation}\label{6}
j_s(z) = \Phi ^+(z) \:\sigma _s \Phi (z),
\end{equation}
where $\sigma _s, \;s=0,1,2,3$ are the Pauli matrices including the identity one $\sigma_0$, they form the basis of any transformations of spinor. Using spinor $\Phi (z)$ (4) and its Hermitian conjugate 
$\Phi ^+ = \Vert\Phi_+^*, \Phi_-^*\Vert$, one has obvious form of observables 
\cite{13,14,8}

\begin{equation}\label{7}
\begin{array}{cc}
j_0 = \Phi_+^* \Phi_+ + \Phi_-^* \Phi_- = a^2 + b^2 ,
&
j_1 = \Phi_+^* \Phi_- + \Phi_-^* \Phi_+ = 2ab \cos (\beta - \alpha) , \\
j_3 = \Phi_+^* \Phi_+ - \Phi_-^* \Phi_- = a^2 - b^2, 
&
j_2 = -i(\Phi_+^* \Phi_- - \Phi_-^* \Phi_+) = 2ab \sin (\beta - \alpha).
\end{array}
\end{equation}

    As far as spinor is defined up to a phase factor, the Hermitian forms are dependent on only three real variables $a, b, (\beta - \alpha)$.

 It is obviously that the Hermitian forms (7) satisfy the identity

\begin{equation}\label{8}
j_0^2 = j_1^2 + j_2^2 + j_3^2,
\end{equation}
containing all of them, it is valid everywhere and under any conditions, therefore it may be considered as the completeness condition for the set of these Hermitian forms.

    Such Hermitian forms may also be constructed on the basis of wave function and its derivative. Taking into account relations (2), one may express the Hermitian forms to be found as following 

\begin{equation}\label{9}
\begin{array}{cc}
j_0 = k\psi \psi ^* + (\psi ')(\psi ^{*'})/k ,
&
j_1 = k\psi \psi ^* - (\psi ')(\psi ^{*'})/k , \\
j_2 = \psi \psi^{*'} + \psi ^* \psi ' ,
&
j_3 = i(\psi \psi^{*'} - \psi ^* \psi ') ,
\end{array}
\end{equation}
they coincide with expressions (7) and satisfy the same identity (8).

     A quantum particle moving under different conditions is described by the Schroedinger equation. Its solution is the complex wave function, therefore it can not be observed directly, and one needs to use the Hermitian forms mentioned above to compare a theory  with experiment. Some of them are conserving due to conservation laws corresponding to the equation, other of them are changing in different processes, and all of them  together form the complete set of observables at any time and in any point.

    Then the question arises: does some equation or its system for observables, i.e. an Hermitian forms, which exclude unobservables and which may be used for description of quantum particle, exist?

     Let us find a differential relations for the set of Hermitian forms (9). Beforehand, differentiating the last expression for $j_3$ and taking into account the Schroedinger equation together with its complex conjugate, one has an ordinary equation for a "probability density current" $j_3' = 0$, or $\nabla \boldsymbol j = 0$ for the stationary partial differential equation
 \cite{6}
which, however, is not considered here.

    Applying the same procedure to the rest expressions in (9), we obtain a set of four equations for the complete set of Hermitian forms

\begin{equation}\label{10}
 j'_0 = \frac{k'}{k} j_1, \quad j'_1 =  2 k j_2 + \frac{k'}{k} j_0,\quad
 j'_2 = - 2 k j_1,\quad  j'_3 = 0.
\end{equation}

    Of course, the same set may be carried out for the Hermitian forms expressed in terms of spinor components, using the spinor representation of the Schroedinger equation (3).

    It is interesting to note that these equations allow one to express  the parameter $k(z)$ in the Schroedinger equation (1) only in terms of observables $j_s$ and their derivatives due to the third equation in (10).
This circumstance may be used to set a geometric content of the parameter $k(z)$ in the Schroedinger equation, and also the same for the parameter $k'/k$ due to the first equation in (10). It would be useful, in turn, for clarification of spatial behaviour of quantum particle described by the complete set of observables.

      It is relevant to put a question: is the set (10) complete, or not? Evidently, its completeness may take place only under      
 the $j_s$ set completeness. Then, differentiating completeness condition (8), one has $j_0 j_0' = j_1 j_1' + j_2 j_2' +j_3 j_3' $. Substituting equations (10) into this expression, one has also an identity. Thus, a use of  the Hermitian forms (6) or (9) leads not only to obvious completeness condition for them. It leads also to the similar condition for its increments. It means that they are consistent for the Schroedinger equation also with external potential, i.e. under constant or changing $k(z)$. More certainly, the set of equations (10), being also complete one as the set of the Hermitian forms (6) or (9), contains all possible conservation laws for observables of the Schroedinger equation under different conditions, if one will put $j_s'=0$ for any $s$. However, investigation of the conservation law  $j_1'=0$ goes out of the paper.

     Let us return to the spinor representation of the Schroedinger equation
(3). The Hermitian forms expressed in terms of wave function and its derivative on the one hand, and those expressed by means of spinor components on the other hand, are the same under connections (2). Furthermore, the equation (3) leads to the same its increments, therefore both approaches lead to the same observables dependence on coordinates and problem parameters. Nevertheless, equation (3), being a pair of first order equations, is more preferable with respect to the Schroedinger equation due to opportunity of groups representations use to investigate a group-theoretic properties of propagators transformations, so as conservation laws.

   Two ways may be used to obtain  spinor representation (3) of  the Schroedinger equation. 

   The first one is a substitution of $\psi$ and $\psi '$ from expression (2) into the Schroedinger equation to obtain pair of first order equations for $\Phi_{\pm}$.

    The second one is connected with physical content of function $k^2(z)$
in the Schroedinger equation (1). Usually this function is supposed to be a difference between kinetic and potential energy of particle. It allows one to use the method
\cite{10,11,12}
 based on division of potential into sequence of small stepwise segments and requirements of $\psi (z)$ and $\psi '(z)$ continuity at common points of neighboring infinitesimal steps. Such procedure leads to the matrix sewing  $\psi$ and $\psi '$ between such small segments continuously, moreover, both of them as a functions of coordinates and also as a function of parameter $k$. Significance of the last circumstance will be discussed below.  Limit of consecutive products of these, almost unit under $\Delta z \to 0$ (and also $\Delta k \to 0$) but non-commutative in general case matrices, leads to the product integral
\cite{15}
introduced by Volterra in 1887. Then one has a solution for spinor 
$\Phi (z_f) = Q(z_f,z_i) \Phi (z_i)$,
where $z_i$ and $z_f$ are initial and final points respectively, and where matrix  $Q(z_f,z_i)$ is expressed as

\begin{equation}\label{11}
Q=\lim_{\substack{N\to\infty \\
\Delta z\to 0}}
\prod_{m=1}^N\exp [ik_m\Delta z_m \sigma _3 + (\Delta k_m /2k_m) \sigma _1] \equiv \\
T \exp \int\limits_{z_i}^{z_f} [ik dz \sigma _3+ \frac{dk}{2k} \sigma _1].
\end{equation}

   An analysis of last expression
\cite{10,11,12,14}
shows that $det Q = 1, Q_{21}=Q_{12}^*, Q_{22}=Q_{11}^*,$ i.e. matrix propagator belongs to the group $SU(1,1).$
  
    It is necessary to emphasize that this propagator leads not only to the spinor components continuity as a function of $z$ and $k$ everywhere due to sewing procedure mentioned above, or the same for $\psi$ and $\psi '$, that leads in turn to continuity of all Hermitian forms, i.e. of all observables. Besides, the propagator (11) belonging to the group $SU(1,1)$ leads, in accordance with the Noether theorems, to fulfillment of all conservation laws for the Schroedinger equation.

     An integrand under sign of the product integral in expression (11) allows geometric interpretation of propagators for the Schroedinger equation in its spinor representation. Being written as matrix two by two in the basis of Pauli matrices including the identity $\sigma_0$, an integrand may be considered as some vector in the space of propagators logarithms. The Pauli matrices may be said to be analogous to the unit vectors of orthogonal basis at the same time
\cite{16,17}.
Taking into account also continuity of the propagator $Q$ from (11) as a function both of coordinates and parameter $k$, a length of vector $ds$ squared of infinitesimal transformation defined by integrand, may be written in form
\cite{13,14,9}

\begin{equation}\label{12}
ds^{2} = -k^{2} dz^{2} + \frac{dk^2}{4k^2}.
\end{equation}

This expression may be considered as the metric of the space of propagators logarithms transformations. Such kind of metric, which is compatible with other its forms accepted in literature and leading to the same Gaussian curvature, means that the propagators logarithms space is the plane with constant negative Gaussian curvature, i.e. the Lobachevsky plane, with $C_G =-4$.  

Besides, it was shown in
\cite{14,9}
that only this  special value of Gaussian curvature
among all, which may have spaces with constant negative Gaussian curvature, leads to the wave equations similar to the Schroedinger or Helmholtz ones. In addition, nonzero Gaussian curvature of this space represents also non-commutativity of transformations of the Schroedinger equation solutions.

     Thus, the space of propagators logarithms of the Schroedinger equation is the Lobachevsky plane with unique Gaussian curvature $C_G = -4.$ It should be noted that possibility of identification of such kind propagators space with the Lobachevsky one is closely connected with isomorphism of the groups $SU(1,1)$ and $SL(2,R)$
\cite{18}.

    Having determined the metric and the Gaussian curvature of the space, one may find an appropriate geometric image for an integrand, and further, for a propagator in the expression (11). Taking into account orthogonality of the Pauli matrices
\cite{17}, 
both terms in integrand may be mapped on the Lobachevsky plane as the oriented orthogonal segments of geodesic lines in accordance with
\cite{13,19}.
Furthermore, one may note that consideration of propagators as a geodesic lines segments in the Lobachevsky space allows one to solve geometric problems of such kind geometry. In turn, it may be found to be useful for physical problems.

    One may note that even if only two of the Pauli matrices entered the integrand  in expressions (11) evidently, the product integral includes all these matrices together with $\sigma_0$ since its expression is product of similar matrices. It means that dimension of the space mentioned above is defined by all Pauli matrices which together with $\sigma_0$ form the basis of all possible transformations described by matrices two by two. For example, a dimension of this space is the same, both in the case of the unidimensional Schroedinger equation or in the case of non-unidimensional one, including the time Schroedinger equation.

     The geometric mapping of matrix propagators into the Lobachevsky space had allowed one to establish the non-Euclidean superposition principle for alternative propagators which takes into account their non-commutativity
\cite{13,19,14,9}. 
It contains four binary compositions of non-commutative matrix propagators, all of them belong to the same group as both entered the compositions and have necessary properties with respect to permutations and inversions, and go to the ordinary Euclidean superposition principle under corresponding conditions. Two of four compositions contain irreversibility, although each of two non-commutative propagators entered these two compositions, is the solutions of the reversible Schroedinger equation 
\cite{20,21}.

    Besides geometric interpretation of propagators, it is extremely important for physical purposes to determine a space of observables $j_s$.

     One may often listen that a particle being described by the unidimensional Schroedinger equation, for example with only 
$z$-dependence of potential (and may be its derivatives) in the equation, would be moved strictly along the same axis. In particular, the authors of
\cite{6}, 
considering the problem on quantum particle moving above unidimensional potential step, supposed conserving "probability density current" 
$j_3 = i(\psi \psi^{*'} - \psi ^* \psi ')$ to be directed along such axis, transmitted and reflected particles are moving along this current, i.e. along
$z$ - axis.

    This point of view seems to be hardly satisfactory. A  quantum particle motion is defined by all Hermitian forms of its complete set which may be constructed on the basis of the Schroedinger equation solutions. This set includes four Hermitian forms, only three of them are independent due to the identity (8). The number of these forms does not depend on if equation is defined by one or more variables. It is defined by the dimensionality of the group, transforming solutions
 \footnote{Since the time Schroedinger equation contains only first order time derivative, it has the same Hermitian forms complete set as the stationary one, i.e. the set (9).}, i.e. $SU(1,1)$ in this case. 

    In particular, it means that the tangent to the line along which all conservation laws are fulfilled  is defined by all Hermitian forms and may be by their derivatives.Therefore, the condition (8) may be considered not only as completeness condition but also as the circumstance that $j_s$ form some vector in the Euclidean space
\cite{8,9}, or, more rigorously, in the space with zero Gaussian curvature.The requirements of $\psi$ and $\psi '$ continuity fulfilled for them as for solutions of the second order differential equation lead to continuity of all Hermitian forms as well. The group-theoretic approach provides fulfillment of necessary conservation laws in accordance with the Noether theorems. Both these circumstances allow one to suppose that a consequence of the points where these conservation laws are fulfilled form the continuous line. This line may be considered as the quantum particle trajectory. 

    If it is not so, then one assumes that a particle may be found at the points where conservation laws had been violated.

    Consideration of observables $j_s$ as orthogonal components of (path) velocity ($j_0$ is its absolute value in such interpretation) in the Euclidean space allows one to attain a second, along with probabilistic, interpretation of the Hermitian forms in quantum mechanics
\cite{8,9}.
Furthermore, connection of $j_s$ and its derivatives with curvature and torsion leads to set the line for point-like object described by the Schroedinger equation, along which all conservation laws are fulfilled, under known of all initial conditions, of course. It is well known from differential geometry
\cite{22}
that these two parameters, the curvature and the torsion, define the spatial line to within a position in space.

     In particular,  free particle under $k=const$ and arbitrary initial conditions is moving along spiral line having the curvature and the torsion to be fixed as far as all necessary conservation laws are fulfilled along this line. The last circumstance leads to an opportunity to consider free quantum particle trajectory as the Euclidean straight line on the Euclidean plane with zero Gaussian curvature which is rolled up into the cylinder surface with the same Gaussian curvature.

    Such behaviour of free quantum particle allows one to propose a qualitative explanation
\cite{8,9} 
double-slit experiment under extremely low intensity of a particles source
\cite{23,24}.

      Obviously, the particle at the potential step may be considered similar to described above but the trajectory in this case would be disposed onto the conical surface with the same zero Gaussian curvature due to the identity (8) as well. The last one is also fulfilled at the step when propagator and corresponding Hermitian forms are varying together with variation of $k$ under conditions $j_2=const, j_3=const$,
\cite{8,9}
as it would be seen below.

\section{Group-theoretic approach to the potential step problem}
\pst

        Let us consider the simplest problem of quantum mechanics, the problem which is supposed to be a standard one and which is solved in
\cite{6}
 - the problem on transition of quantum particle above potential step.

      Let a potential step is located at the point $z_1$, positive parameter $k(z)$ in the Schroedinger equation is changed as

\begin{equation}\label{13}
k(z) =
\left \{
 \begin{array}{l}
k_1 = const, \quad if \quad z < z_1,\\
k_2 = const, \quad if \quad z > z_1,
 \end{array}
\right.
\end{equation}
and quantum particle starting at $z_0 < z_1$ goes from the left to the right, passing through a step-like potential at $z_1$ above it, i.e. $k(z)>0$ everywhere.

    We shall also accept that the initial conditions at the point $z_0$ are defined in most general form as in equalities (5): 

\begin{equation}\label{14}
\Phi_+(z_0) = a_0 e^{i\alpha _0}, \quad
\Phi_-(z_0) = b_0 e^{i\beta _0}. 
\end{equation}{}

    Keeping in mind subsequent comparison of the problem solution here with the same in
\cite{6},
we also write the corresponding wave function and its derivative at the same point $z_0$, expressing them from equalities (2):

\begin{equation}\label{15}
\psi (z_0) = \frac{k_1^{-1/2}}{\sqrt 2}
(a_0 e^{i\alpha_0} + b_0 e^{i\beta_0}),\quad
\psi ' (z_0) = i\frac{k_1^{1/2}}{\sqrt 2}
(a_0 e^{i\alpha_0} - b_0 e^{i\beta_0}).
\end{equation}

    The Hermitian forms at the same point $z_0$ are the following

\begin{equation}\label{16}
\begin{array}{cc}
j_0(z_0) = a_0^2 + b_0^2 ,
&
j_1(z_0) = 2a_0b_0 \cos (\beta_0 - \alpha_0) , \\
\\
j_3(z_0) =  a_0^2 - b_0^2, 
&
j_2(z_0) = 2a_0b_0 \sin (\beta_0 - \alpha_0).
\end{array}
\end{equation}

    Obviously,  the identity (8) is fulfilled under these Hermitian forms.

    Let us calculate a propagators in each area along axis $z$ in accordance with
\cite{8,9}.

    The parameter $k(z) = k_1$ is constant  everywhere in $z_0\le z < z_1,$  $k'=0$, an integrand is commutative everywhere in this area, therefore

\begin{equation}\label{17}
Q(z,z_0) = T\exp\int\limits_{z_0}^{z}
 ik_1 dz \sigma _3 = 
e^{ik_1(z-z_0)\sigma_3}  \equiv e^{iN\sigma_3},\quad z_0\le z < z_1.
\end{equation}{}

     As far as
 $Q^+ \sigma_3 Q = \sigma_3$ and  $Q^+ \sigma_0 Q = \sigma_0$,  
then $j_3(z)=j_3(z_0), j_0(z)=j_0(z_0)$, i.e.
 the Hermitian forms $j_3$ and $j_0$ from (6) are constant everywhere in this area, and two others vary.

    The matrix of product integral in $\varepsilon$-vicinity of $z_1$ under 
$\varepsilon \to 0$ is following

\begin{equation}\label{18}
Q(z_0+\varepsilon,z_0-\varepsilon) =
T\exp\int\limits_{z_0-\varepsilon}^{z_0+\varepsilon}
\frac{k'}{2k} dz \sigma _1 = 
 T\exp\int\limits_{k_1}^{k_2}
\frac{dk}{2k}\sigma _1 = 
e^{\frac{1}{2}\ln \frac{k_2}{k_1}\sigma_1}\equiv e^{L\sigma_1}.
\end{equation}{}
   This matrix describes interaction with a potential step due to non-diagonal
matrix $\sigma_1$.

   As far as 
 $Q^+ \sigma_3 Q = \sigma_3$ and  $Q^+ \sigma_2 Q = \sigma_2$,  
then the Hermitian forms $j_3$ and $j_2$ are constant on the step, and two others are varied
\cite{8}.

    A propagator for $z>z_1$ is calculated similar to the same for $z<z_1$, the result is following:

\begin{equation}\label{19}
Q(z,z_1) = T\exp\int\limits_{z_1}^{z}
 ik_2 dz \sigma _3 = 
e^{ik_2(z-z_1)\sigma_3}  \equiv e^{iM\sigma_3},\quad z> z_1,
\end{equation}{}
and the same Hermitian forms are conserved.

   Now we can write solutions and Hermitian forms at different areas along axis $z$.

    The spinor $\Phi (z)$ at arbitrary point $z$ under $z_0 \leq z<z_1$,   $\Phi (z) = Q(z,z_0)\Phi (z_0)$ with $Q$ from (17), leads to two spinor components 

\begin{equation}\label{20}
\Phi_+(z) = e^{ik_1(z-z_0)} a_0 e^{i\alpha _0},\quad 
\Phi_-(z) = e^{-ik_1(z-z_0)} b_0 e^{i\beta _0}.
\end{equation}

    The wave function and its derivative are following there:

\begin{equation}\label{21}
\begin{array}{c}
\psi (z) = \frac{k_1^{-1/2}}{\sqrt 2}
\{e^{ik_1(z-z_0)} a_0 e^{i\alpha _0} +
 e^{-ik_1(z-z_0)} b_0 e^{i\beta _0}\},\\
\psi '(z) = i\frac{k_1^{1/2}}{\sqrt 2}
\{e^{ik_1(z-z_0)} a_0 e^{i\alpha _0} -
 e^{-ik_1(z-z_0)} b_0 e^{i\beta _0}\}.
\end{array}
\end{equation}

    Corresponding Hermitian forms may be calculated by means of equalities (7) expressing them in terms of ${\Phi_\pm}$ from (20)

\begin{equation}\label{22}
\begin{array}{cc}
j_0(z) = a_0^2 + b_0^2 ,
&
j_1(z) = \cos [2k_1(z-z_0)]\cdot j_1(z_0) + 
\sin[2k_1(z-z_0)]\cdot j_2(z_0), \\
j_3(z) =  a_0^2 - b_0^2, 
&
j_2(z) = -\sin [2k_1(z-z_0)]\cdot j_1(z_0) +
\cos[2k_1(z-z_0)]\cdot j_2(z_0).
\end{array}
\end{equation}

    What is influence of the potential step on observables? Let us calculate the propagator including the step to answer the question. It is $Q(z_1+\varepsilon, z_0)$ under $\varepsilon \to 0$, i.e. the propagator to be defined is the product of propagators from (17) and
 (18):\: $\exp({L\sigma_1})\cdot \exp({iN\sigma_3})$. It leads to the spinor components immediately behind the step

\begin{equation}\label{23}
\begin{array}{c}
\Phi_+(z_1) =\cosh Le^{ik_1(z_1-z_0)} a_0 e^{i\alpha _0} +
\sinh L e^{-ik_1(z_1-z_0)} b_0 e^{i\beta _0},\\
\Phi_-(z_1) =\sinh Le^{ik_1(z_1-z_0)} a_0 e^{i\alpha _0} +
\cosh L e^{-ik_1(z_1-z_0)} b_0 e^{i\beta _0}.
\end{array}
\end{equation}
    Here we do not  write wave function and its derivative, they may be found by means of these spinor components and equalities (2).

    Calculations of the observables immediately behind the step
(under $\varepsilon \to 0$)
 accordingly to (7) with spinor components from (23) lead to

\begin{equation}\label{24}
\begin{array}{c}
j_0(z_1+\varepsilon) =\cosh2L\cdot j_0(z_1-\varepsilon) +
 \sinh2L\cdot j_1(z_1-\varepsilon)=\\
\\
\cosh2L\cdot j_0(z_0) + \sinh2L \{\cos[2k_1(z_1-z_0)]\cdot j_1(z_0) +
\sin[2k_1(z_1-z_0)]\cdot j_2(z_0)\},\\
\\
j_1(z_1+\varepsilon) =\sinh2L\cdot j_0(z_1-\varepsilon) + 
\cosh2L\cdot j_1(z_1-\varepsilon) =\\
\\
\sinh2L\cdot j_0(z_0) + \cosh2L \{\cos[2k_1(z_1-z_0)]\cdot j_1(z_0) +
\sin[2k_1(z_1-z_0)]\cdot j_2(z_0)\},\\
\\
j_2(z_1+\varepsilon) = j_2(z_1-\varepsilon) =\\
\\
-\sin[2k_1(z_1-z_0)]\cdot j_1(z_0) + \cos[2k_1(z_1-z_0)]\cdot j_2(z_0),\\
\\
j_3(z_1+\varepsilon) = j_3(z_1-\varepsilon) = j_3(z_0).
\end{array}
\end{equation}

     Now one needs to define a wave function, its derivative and observables everywhere behind the step. The propagator there is a consecutive product  
of three ones from (17-19): 
$ \exp({iM\sigma_3})\cdot \exp({L\sigma_1})\cdot \exp({iN\sigma_3})$.
These calculations lead to two spinor components expressed via  arbitrary initial conditions at $z_0$, distance up to the potential step 
 $(z_1-z_0)$, values of parameters in different areas of free particle motion,
$k_1$ under $z<z_1$ and $k_2$ under $z>z_1$, 
a step of potential described by parameter $L=(1/2)\ln(k_2/k_1)$, and distance up to an arbitrary point $z$: 

\begin{equation}\label{25}
\begin{array}{c}
\Phi_+(z) = e^{ik_2(z-z_1)}[\cosh Le^{ik_1(z_1-z_0)} a_0 e^{i\alpha _0} +\sinh L e^{-ik_1(z_1-z_0)} b_0 e^{i\beta _0}],\\
\Phi_-(z) = e^{-ik_2(z-z_1)}[\sinh Le^{ik_1(z_1-z_0)} a_0 e^{i\alpha _0} +
\cosh L e^{-ik_1(z_1-z_0)} b_0 e^{i\beta _0}].
\end{array}
\end{equation}
      Corresponding expressions for the wave function and its derivative there are following:

\begin{equation}\label{26}
\begin{array}{c}
\psi (z) = \frac{k_2^{-1/2}}{\sqrt 2}
\{ e^{ik_2(z-z_1)}[\cosh Le^{ik_1(z_1-z_0)} a_0 e^{i\alpha _0} +\sinh L e^{-ik_1(z_1-z_0)} b_0 e^{i\beta _0}] +\\
\\
e^{-ik_2(z-z_1)}[\sinh Le^{ik_1(z_1-z_0)} a_0 e^{i\alpha _0} +
\cosh L e^{-ik_1(z_1-z_0)} b_0 e^{i\beta _0}]\},\\
\\
\psi ' (z) = i\frac{k_2^{1/2}}{\sqrt 2}
\{ e^{ik_2(z-z_1)}[\cosh Le^{ik_1(z_1-z_0)} a_0 e^{i\alpha _0} +\sinh L e^{-ik_1(z_1-z_0)} b_0 e^{i\beta _0}] -\\
\\
e^{-ik_2(z-z_1)}[\sinh Le^{ik_1(z_1-z_0)} a_0 e^{i\alpha _0} +
\cosh L e^{-ik_1(z_1-z_0)} b_0 e^{i\beta _0}]\}.
\end{array}
\end{equation}

     Now using the expressions either for two spinor components $\Phi\pm(z)$ from (25) or  for the wave function and its derivative from (26) at any point $z$ behind the potential step under $z>z_1$,
one can express all observables of its complete set  at point $z$ via the same ones at the point $z_0$ in obvious form as

\begin{equation}\label{27}
\begin{array}{c}
j_0(z) = j_0(z_1+\varepsilon) =\\ \cosh2L\cdot j_0(z_0) +
\sinh2L \cos[2k_1(z_1-z_0)] \cdot j_1(z_0) + 
\sinh2L \sin[2k_1(z_1-z_0)] \cdot j_2(z_0),\\
\end{array}
\end{equation}

\begin{equation}\label{28}
\begin{array}{c}
j_1(z) = \cos[2k_2(z-z_1)] \sinh2L \cdot j_0(z_0) +\\
\\
\{\cos[2k_2(z-z_1)]\cosh2L \cos[2k_1(z_1-z_0)] - 
\sin[2k_2(z-z_1)] \sin[2k_1(z_1-z_0)]\}\cdot j_1(z_0) +\\
\\
\{\cos[2k_2(z-z_1)]\cosh2L \sin[2k_1(z_1-z_0)] + 
\sin[2k_2(z-z_1)] \cos[2k_1(z_1-z_0)]\}\cdot j_2(z_0),
\end{array}
\end{equation}

\begin{equation}\label{29}
\begin{array}{c}
j_2(z) = - \sin[2k_2(z-z_1)] \sinh2L \cdot j_0(z_0) +\\
\\
\{-\sin[2k_2(z-z_1)]\cosh2L \cos[2k_1(z_1-z_0)] - 
\cos[2k_2(z-z_1)] \sin[2k_1(z_1-z_0)]\}\cdot j_1(z_0) +\\
\\
\{-\sin[2k_2(z-z_1)]\cosh2L \sin[2k_1(z_1-z_0)] + 
\cos[2k_2(z-z_1)] \cos[2k_1(z_1-z_0)]\}\cdot j_2(z_0),
\end{array}
\end{equation}

\begin{equation}\label{30}
j_3(z) = j_3(z_0).
\end{equation}

    Obviously that all observables are continuous not only as a function of $z$ but also as a function of parameters $k_1$ and $k_2$. In particular, all observables go over to these ones for free particle: formulae (27)-(30) go to formulae (22) under condition $k_2 \to k_1$. 

    It is relevant to put a question: may one describe quantum particle being subjected to the Schroedinger equation by means of only observables? Such approach was formulated by Heisenberg at the beginning of quantum mechanics.

    To make it more clear let us consider the complete system of differential equations (10) for the complete set of Hermitian forms (6) or (9). 

    At first, we shall restrict with the area of constant $k$, particularly 
$z_0 \leq z < z_1$. The set (10) goes to 

\begin{equation}\label{31}
 j'_0 = 0, \quad j'_1 =  2 k j_2,\quad
 j'_2 = - 2 k j_1,\quad  j'_3 = 0
\end{equation}
under initial conditions at $z_0$ (16). It means that the conservation law $j_0=const$ is added to the general conservation law for the stationary Schroedinger equation $j_3=const$ in this area. The pair of equations for $j_1$ and $j_2$ forms a closed system, it may be solved under initial conditions for them. One also sees from (31)  that
$j_1^2+j_2^2 =const$, it is consistent with $j_0=const, j_3=const$ and the identity (8).

     Now one may consider the potential step itself. Let us multiply all equations (10) by small interval $\Delta z$ which includes the point $z_1$, then they are transformed into the following set

\begin{equation}\label{32}
\Delta j_0 = \frac{\Delta k }{k}j_1, \quad 
\Delta j_1 =  2 k  \Delta z j_2 + \frac{\Delta k}{k}j_0,\quad
\Delta j_2 = - 2 k \Delta z j_1,\quad 
\Delta j_3 = 0,
\end{equation}
which under restricted $k$ and $\Delta z \to 0$, go over to two conservation laws  at the step: 
 $j_2=const, j_3=const$, and to pair of equations

\begin{equation}\label{33}
\frac{dj_0}{dk} = \frac{j_1}{k},\quad \frac{dj_1}{dk} = \frac{j_0}{k}.
\end{equation}
     The last one is a closed system of two ordinary differential equations for $j_0$ and $j_1$ as a function of $k$ with conditions for them at $z_1$
$j_0(k_1)$ and $j_1(k_1)$ immediately before the potential step.

    To solve this system, let us go over to the second order equation for one of $j_0$ or $j_1$, for example $k^2j''_0+k j'_0-j_0=0,$ and find its solutions
as $k$ to power $\lambda$, then $\lambda=\pm 1$. It leads to the solutions for $j_0(k)$ and $j_1(k)$ on the potential step as

\begin{equation}\label{34}
\begin{array}{c}
j_0(k) = \frac{1}{2}(\frac{k}{k_1}+\frac{k_1}{k}) j_0(z_1-\varepsilon)+
\frac{1}{2}(\frac{k}{k_1}-\frac{k_1}{k}) j_1(z_1-\varepsilon),\\
j_1(k) = \frac{1}{2}(\frac{k}{k_1}-\frac{k_1}{k}) j_0(z_1-\varepsilon)+
\frac{1}{2}(\frac{k}{k_1}+\frac{k_1}{k}) j_1(z_1-\varepsilon),\\
\end{array}
\end{equation}
where
$j_0(z_1-\varepsilon), j_1(z_1-\varepsilon)$
are the corresponding values immediately before the step under $\varepsilon \to 0 $.These solutions go over to equations (24) immediately after the potential step under $L=(1/2)\ln (k_2/k_1)$.

    Of course, the identity (8) is fulfilled both under $k(z)=const$ and also on the potential step.

    The area $z>z_1$ under $k(z)=k_2=const$ may be considered similar to the area $z<z_1$ but with $k_2=const$.

     Thus, at least in the simplest case of the potential step, an opportunity  to calculate observables for quantum particle as a solution of differential equations for them exists. 

It seems to be evident that similar conclusion can not be carried out when some kinds of compositions of wave functions or spinors are necessary, in particular under use of some compositions of the non-Euclidean superposition principle
\cite{13,19,9}.
It is more then doubtful, taking in account the Euclidean character of the identity (8),  that the system (10) contains the peculiarities connected with the Gaussian curvature $C_G = -4$ of the solutions transformations space, which is implicitly included into the Schroedinger equation or its spinor representation.

\section{Classical solution of the potential step problem}
\pst

       Now it is appropriate to quote calculations on the same problem according to the classical approach generally accepted in quantum mechanics. We shall do it very near to
\cite{6}, 
but including some negligible details excluded by the authors but which, however, arise from their consideration and  which are desirable to achieve more clarity.

     Thus, let us suppose that quantum particle is described by unidimensional stationary Schroedinger equation as if it would be moving from left to right along axis $z$ in the direction of the potential step at the point $z=z_1$. Everywhere on the left side, before the step, a solution of the Schroedinger equation is supposed to be presented accordingly to the authors as

\begin{equation}\label{35}
\psi (z) = e^{ik_1(z-z_1)} + r e^{i\gamma}e^{-ik_1(z-z_1)}, \;z\leq z_1,
\end{equation}
with $k_1 =const, r $ and $\gamma$ are real constant values to be defined. The first term in (35)  presents an incident wave with normalized amplitude, the second one corresponds to a reflected wave with unknown amplitude and phase. The expression for $\psi$-function before the step is chosen to be near to one from
\cite{6},
 and it may be consistent with its expression from (21) under $a_0=\sqrt{2k_1}, \alpha_0=-k_1(z_1-z_0), b_0=\sqrt {2k_1}r$, and $\beta_0=\gamma +k_1(z_1-z_0)$. It means that both expressions for $\psi (z)$ under $z<z_1$, i.e. (35) (accordingly to
\cite{6}) 
and (21), are chosen in quite general form and may be reduced one to another in any case.

   Everywhere on right side, behind the step, a solution is presented by the authors as following

\begin{equation}\label{36}
\psi (z) =  t e^{i\delta}e^{ik_2(z-z_1)}, \;z \ge z_1,
\end{equation}
where $k_2=const, t$ and $\delta$ are amplitude and phase of  the transmitted wave, they are also unknown real fixed parameters.

    One may see from the expressions (35) and (36) that the wave functions are separately continuous in both areas, on the left side and on the right side of the step, both as functions of coordinate $z$ and, also separately, of parameters $k_1$ and $k_2$. Therefore the wave function is differentiable along $z$ on the left side and on the right side of the potential step up to the point $z_1$, and one may require at least the wave function and its derivative continuity also at $z_1$ as functions of $z$. These two requirements had been used by the authors to find parameters defining quantum particle motion, they lead in turn to two expressions for complex coefficients 

\begin{equation}\label{37}
t e^{i\delta} = \frac{2k_1}{k_1 + k_2},\quad
r e^{i\gamma} = \frac{k_1 - k_2}{k_1 + k_2},
\end{equation}
describing, accordingly to the authors, transmitted and reflected waves 
respectively.

Then the authors,  omitting phases calculations, go over up to amplitudes of transmitted and reflected waves with the aim of calculation transmission and reflection coefficients. 

    It had been noted in
\cite{6}
 that currents of the "probability density" in incident, reflected and transmitted waves are proportional to $k_1$,  
$k_1 \vert r e^{i\gamma}\vert^2 = k_1 r^2 $ and 
$k_2 \vert t e^{i\delta}\vert^2 = k_2 t^2 $ respectively. Defining reflection $R$ and transmission $T$ coefficients as ratio of corresponding currents to the incident one, they obtain $R=r^2$ and $T=(k_2/k_1)t^2$.

Taking into account expressions  
\begin{equation}\label{38}
t^2 =  \vert t e^{i\delta}\vert^2 = \frac{4k_1^2}{(k_1 + k_2)^2},\quad r^2 = \vert r e^{i\gamma}\vert^2 = \frac{(k_1 - k_2)^2}{(k_1 + k_2)^2},
\end{equation}
it leads to the identity $R +T = 1$ for the reflection and transmission coefficients as if it would be the conservation law.

\section{Comparison of two approaches}
\pst

      The common initial point for comparison of two approaches described above is the stationary unidimensional Schroedinger equation applied to the simplest problem of quantum mechanics. Independently of a theory or a method of the problem solution, it seems to be appropriate first of all to compare those results which are closely connected with an experiment. Therefore we shall compare the observables, i.e. the Hermitian forms, which had been obtained for the problem under the group-theoretic approach, on the one hand, and those which had been obtained in 
\cite{6}
and also those which may be calculated within the framework of their solution, on the other hand. Then we shall attempt to set causes of distinctions of the results achieved in two approaches.

"Nobody understands quantum mechanics", but everybody can obtain all Hermitian forms constructed on the base of the Schroedinger equation solutions. It is more so as far as the similar constructions are well known under the name of the Stokes parameters during more then hundred years. Their main properties had been described, for example,  by M.Born
\cite{25} 
who is one of the founders of the probability interpretation in quantum mechanics at the same time.

    It is quite obviously that only four Hermitian forms may be constructed on the base of the equation (1) complex solutions and their first derivatives for any point $z$, including the initial one $z_0$. They are following
$$
\psi \psi^*, \psi ' \psi^{*'}, \psi \psi^{*'} + \psi ' \psi^*,
i(\psi \psi^{*'} - \psi ' \psi^*).
$$

The first and the last of these Hermitian forms are universally recognized and used in contemporary schemes of quantum mechanics under  names of the "probability density"
 $\rho = \psi  \psi^{*}$ and its current.

    The third Hermitian form is linear with respect to derivative in just the same way as the last one, in addition it coincides with $\rho'$, or with $\nabla \rho$ in the spatial case. Therefore the third Hermitian form is also the current, but the diffusion one, in opposite to the last one which may be named as the convective current \footnote{For example, the currents of similar kind are used for description of particles transport in the nuclear fusion.}. It would be emphasized that the "probability density" current accepted in the existing schemes of quantum mechanics depends on a wave function amplitude and its {\it phase derivative}, whereas a current $j_2=\rho '$ is defined only by its amplitude and its {\it amplitude derivative}, but does not completely depend on its phase. Of course, it is no wonder that two different currents may exist under complex solutions of the Schroedinger equation, and restriction with only one current is quite insufficient in general case. 

     The second Hermitian form, $\psi ' \psi^{*'}$, is bilinear with respect to the wave function derivative in just the same way as the first one, i.e. "probability density" $\psi \psi^*$, is bilinear with respect to $\psi$. Taking into account equal necessity of both of them, i.e. $\psi$ and $\psi'$, at the initial point $z_0$ for solution of the equation (1), one needs to recognize that the composition  $\psi ' \psi^{*'}$ has to be included into the set of  Hermitian forms if $\psi \psi^*$ is already included into consideration. It is especially evident under spinor representation of the Schroedinger equation, both spinor components are equivalent there, and they are included into the complete set of the Hermitian forms (6) under the same rights. 

      All Hermitian forms which had been obtained in accordance with the group-theoretic approach and which correspond to three areas of potential are reduced in the third part of the paper. However the first pair of the Hermitian forms mentioned above is included there as its linear combinations, i.e.\\ 
$\psi \psi^* = (j_0+j_1)/(2k), \psi ' \psi^{*'}  = k(j_0-j_1)/2$.

   Now it should be brought all Hermitian forms calculated everywhere along $z$ on the base of the expressions for wave functions reduced according to
\cite{6}.

    In spite of the authors had not calculated all of them, they had used  continuous wave functions (35) and (36), and moreover they had used the continuity requirements for them and its derivatives, in particular at the potential step. Therefore one may calculate all Hermitian forms which had to be also continuous. Using the wave function (35), one has the following expressions for the Hermitian forms everywhere before the potential step, i.e. under 
$z < z_1$:

\begin{equation}\label{39}
\begin{array}{cc}
j_0 = 2k_1(1+ r^2) ,
&
j_1 = 4k_1 r \cos[2k_1(z-z_1)-\gamma] , \\
j_2 = -4k_1 r \sin[2k_1(z-z_1)-\gamma] ,
&
j_3 = 2k_1(1- r^2).
\end{array} 
\end{equation}

    It is seen from the expressions (39) that all Hermitian forms differ from zero in general case, and the identity (8) is fulfilled at the same time. It is
interesting to note that derivative of the "probability density" considered in the general accepted forms of quantum mechanics,
 $\rho=\psi \psi^*=(j_0+j_1)/(2k)$, is equal to the unconsidered diffusion current of the "probability density", $\rho ' = j_2$, also within the framework of the problem solution in
\cite{6}.

    The Hermitian forms calculated by means of the wave function (36) for
$z > z_1$ are following:

\begin{equation}\label{40}
j_0 = 2k_2 t^2,\quad j_1 = 0,\quad j_2 =0,\quad j_3 = 2k_2 t^2. 
\end{equation}

   All these Hermitian forms are constant everywhere, and the identity (8) is also fulfilled.

     The wave function (35) has highly general form, and it is continuous together with derivative. Therefore the initial conditions similar to (15) may be also set, and one has opportunity to consider this problem also as the initial one to compare with the stated in section 3. Such approach allows one to use the identity (8) which can not be used in the approach from
\cite{6}.

   Obviously that the Hermitian forms (39) and (22) in the area $z < z_1$ are almost the same up to notations. However, the expressions (22) do not contain any properties of the potential step, whereas expressions (39) contain them. This circumstance creates an impression as if potential step, which may be arbitrary, influences on the initial conditions, which may be also arbitrary. A cause of this strange "interaction" between arbitrary step and initial conditions is a choice of insufficiently general expression for wave function (36) behind a step.

    Comparing corresponding expressions for the area behind the step, (40) and (27)-(30) for $z > z_1$, one discovers an acute distinctions. 

    The first and most impressive distinction is that the Hermitian forms $j_1$ and $j_2$ are equivalent to zero everywhere under $z>z_1$  independently on any parameters. In particular it means that even if no matter how small a potential step is, it transforms  both these parameters to zero even no matter how great  they would be before the step. 
  The perturbation theory is not applied in this consideration of the problem even under infinitesimal potential step, when small step has to lead to small variations of Hermitian forms, in particular $j_1$ and $j_2$. One may conclude that, in spite of the wave function and its derivative are continuous everywhere along $z$ including the step, some of the Hermitian forms are discontinuous with respect to small variations of
$\Delta k = k_2 - k_1$: arbitrary $j_1$ and $j_2$ are transformed into zero
ones by a potential step $\Delta k = \pm \varepsilon$ independently on 
$\varepsilon \to 0$, but they conserve the value $j_1^2 + j_2^2$ under
$\varepsilon = 0$, i.e. $k_1=k_2$.

    The second distinction is connected with phases in the coefficients before reflected and transmitted waves. It is obviously from two expressions (37) that $\delta = 0$ under any $k_1$ and $k_2$, but 
$\gamma = 0$ if $k_1>k_2$, and $\gamma = \pi$ if $k_1<k_2$. It is also evidently that these values of phases are the consequence only of pure mathematical requirements to real and imaginary parts of complex equalities (37), and they do not contain any physical causes in the problem considered here. 

   Probably, this circumstance may be considered as display of distinctions between two ways to set a connection of complex values arising in theory with real observables. One of these ways is to accept its real and imaginary parts as observables, another one is to use an Hermitian forms. It seems that the first way has no local completeness condition, particular at the initial point. The second way leads to the local completeness condition (8)
applied also at the initial point.

   The third peculiarity is following. Setting of the problem as in
\cite{6},
 similar to the boundary one, when asymptotic behaviour
of the wave function is accepted in forms (35) and (36), does not allow to use the identity (8), in particular at the initial point, since the value of first order derivative is absent. Four real variables define the problem considered here, being set both as an initial and also as a boundary-value problem. Only three of them are independent due to the identity (8). 
One may construct all Hermitian forms in the first case, the identity (8) inclusive. However, it is impossible in the second case as far as wave function derivatives are not set at both direction of the $z$-axis, on the one hand. On the other hand, it is unknown some kind of the completeness condition containing two different points (or directions, as in
\cite{6}) 
of the same axis
\footnote{Apparently, setting of quantum problem as a boundary one does not allow a use of completeness condition similar to (8).}.
 As a consequence, one has four independent real parameters instead of three, which define the problem. It is relevant to mention here that the group of spinor transformations, essential in quantum mechanics, for example $SU(1,1)$ or $SU(2)$, is the three-parameter group.

     Thus, one may see from the expressions (22), (24) and (27)-(30) that all Hermitian forms of the complete set, calculated on the base of group-theoretic approach, are continuous everywhere and
$j_1$ and $j_2$
oscillate with periods defined by  $k_1, k_2$ in corresponding areas. 
It is also evidently from expressions (39), (40) obtained on the base of
\cite{6}
that both these Hermitian forms oscillate with the same period before step, but all oscillations come to an end  behind it, moreover both of them are zero there independently of step value.

    In other words, if one will set the problem as an initial-value one and will remain within the framework of the method
\cite{6}, 
there is singularity in the vicinity of $k_1=k_2$: oscillating character of motion everywhere under fulfillment of this condition (free particle) but
$j_1 \neq 0, j_2 \neq 0 \to j_1=0, j_2=0$ at the step in any small vicinity outside of $k_1=k_2$. This singularity is absent under the group-theoretic method as it is evident from expressions (27)-(30).

    Consideration in 
\cite{6}
does not offer some kind of geometric interpretation  neither $j_1,
 j_2$ nor, all the more, the complete set of Hermitian forms  neither in the Euclidean space, nor in some another one. Instead of this the authors offer 
geometric interpretation for complex solution as a plane wave with infinite
front as geometric contents of solutions (35), (36).  

    On the contrary, the group-theoretic approach does not contain a geometric interpretation of the complex wave function or spinor components in the Euclidean space. Taking into account noncommutativity of  transformations in quantum mechanics, it would be absurd to look for its geometric content in flat spaces. Instead of it the approach leads to the geometric interpretation of the complete set of real Hermitian forms in the Euclidean space
 \cite{8,9}, or, more exactly, in a space with zero Gaussian curvature.      

    Let us note that the concept of the plane wave leads in turn to the concept of the "collapse of wave function", which seems to be necessary for explanation of the experiments with low intensity particles source
\cite{23,24}, as it is supposed in the generally accepted schemes of quantum mechanics. 

     Existence of the completeness condition (8), connection of $j_s$ with path velocity in the Euclidean space, connection of them and its derivatives with the curvature and torsion defining the line along which all conservation laws are fulfilled, lead to conclusion that free quantum particle moves along helical line in general case, the particle trajectory is settled onto the cylinder surface. Radius and pitch of screw are proportional to the de Broglie wave length of particle
\cite{8}.
 It should be noted that part of trajectory corresponding to the step is settled onto the conical surface, as it was mentioned earlier. Such character of the quantum particle motion allows one to exclude the concept of the "wave function collapse" and to propose qualitative explanation
\cite{8,9}
 of the experiment with the low intensity particle source
\cite{23,24} 
at the same time.

    Here it is relevant to put the question if the asymptotic expression of the wave function in the form of (36) may exist, and under what conditions. Let us consider the question from the group-theoretic
view-point.

    Let the initial conditions had been taken in a general form (14) or (15). Only the potential step may change a particle motion, therefore it is clear that it is enough to investigate only spinor and the Hermitian forms transformations directly on the potential step. We shall use the expressions (23) requiring there $z_1=z_0$. One may see from the first expression in (21) that the required kind of the wave function would be obtained if one of two spinor components is equal to zero. For example, let $\Phi_{-} = 0$ is fulfilled in the last expression in (23). Then one has the condition 
$\tanh L = -(b_0/a_0)\exp{[i(\beta_0-\alpha_0)]}$.
Since $L$ is real, then $(\beta_0-\alpha_0) = 0, \pi$. The last condition is the necessary requirement to obtain $\Phi_{-} = 0$, it is also the condition in order $j_2$ to become equal to zero {\it immediately before} the step. Since 
$L=(1/2)\ln (k_2/k_1)$ the above-mentioned condition is transformed into 
$\pm (k_2-k_1)/(k_2+k_1)=b_0/a_0$, where sign has to be made consistent with sign of $(k_2-k_1)$. The last condition is the requirement for $j_1$ to become equal to zero {\it immediately behind} the step. As a result, the wave function everywhere behind the step will be transformed into the same form as in
\cite{6}.
Evidently, it is the result of interaction of the quantum particle under arbitrary conditions at the initial point with the potential step, but only under special conditions established above. 

    The same result may be obtained in another way. Let us ask yourself if
 the helical trajectory may be transformed into the Euclidean straight line by the potential step. We shall consider the question restricting with only the Hermitian forms, and equations for them.

   It is necessary to obtain $j_1=j_2=0$ everywhere behind the step. In accordance with equations (32), the observable $j_2$ is insensitive to potential step as long as $\Delta z \to 0$. It means that if $j_2=0$ immediately before the step due to 
 $(\beta-\alpha) = 0, \pi$,
 then it would be also zero immediately behind it. Further, requiring $j_1=0$ behind the step in the second equation in (34) under assuming $k=k_2$ there, one obtains the condition 
$$
\frac{k_2^2-k_1^2}{k_2^2+k_1^2}=
 -\frac{j_1(z_1-\varepsilon)}{j_0(z_1-\varepsilon)},
$$
which defines transformation $j_1(z_1+\varepsilon) \to 0$. Then 
$j_1^2(z)+j_2^2(z)=0$ everywhere behind the potential step, and the line along which all conservation laws are fulfilled, i.e. particle trajectory, is the Euclidean straight line.

      Besides, this example shows also a role of phase, or, more exactly, phase difference between two spinor components, or wave function and its derivative, in transformations of the observables in quantum mechanics.

       Let us call attention to the wave functions (35) and (36) once more.

    The authors consider the terms $e^{\pm ikz}$ with correspondent coefficients as the plane waves describing quantum particle
going along $z$-axis. It is assumed that different signs in the exponentials correspond to the waves going (in just the same way as the particle, in accordance with these expressions) in opposite directions along the same axis. Constant parameter $k$ ($k_1$ before the potential step and $k_2$ behind it) is connected with linear momentum (and energy) of particle, and in turn with the de Broglie wave length. It may be assumed that the last circumstance has to lead to spatial periodical variations at least of something which may be measured.

     In fact, the expressions mentioned above, in particular under condition of potential step absence,  $k_1=k_2=k$, i.e.  $\psi=e^{ikz}$ everywhere infinity inclusive, contain periodicity, and any theoretician will say that, of course, there is periodicity with the de Broglie wave length.

     Let us calculate only the Hermitian forms which are accepted in general recognized schemes of quantum mechanics. Then one has the probability density $\rho=\psi \psi^*=1$ and its current 
$j=i(\psi \psi^{*'} - \psi ' \psi^*)=2k$ to be constant everywhere infinity inclusive. The periodicity with this period is disappeared, and any experimentalist may verify it by measurements.

   Here we have an acute contradiction between the theoretical concept of plane wave together with its probability interpretation, and experimental data. Requirements of $\delta$-normalization with respect to any parameter do not change this situation - this procedure can not influence on existence or absence of periodicity.

     It is relevant to compare these conclusions with the results obtained within the framework of the group-theoretic approach: the expressions (22) as long as (27)-(30) under absence of potential step show periodicity for some observables. Inclusion of potential step leads to similar periodicity, but with different wavelengths in different areas of potential. Periodicity is appeared as particle motion along helical lines. It should be noted that the helical line radius and pitch of a screw also vary together with $k$
\cite{8}.

    It is necessary to make remark. The group-theoretic approach partly realized here for the example of quantum particle transmission 
above the potential step allows one to obtain the spatial lines in the Euclidean space along which all necessary conservation laws are fulfilled. But the approach does not allow to obtain the reflection and transmission coefficients. One may see that the situation is vice versa with respect to one in
\cite{6},
where these coefficients are defined and obtained, but where these lines are not determined.

    Both coefficients are defined under confidence that the Euclidean superposition principle is applicable in quantum mechanics almost without restrictions. It would be correct under the condition of rejection of tracing the theory as the group-theoretic one, and in turn under rejection of the Noether theorems as a consequence, of course.

   In general, it seems that such defined coefficients, in just the same way as a scattering amplitudes, are inconsistent with the group-theoretic scheme due to introduction of the second binary operation. Definition of the scattering amplitude is based on addition of incident and reflected, or scattered, waves.

 Nevertheless, the quantum scattering theory, just as the phase functions method
\cite{26}
 developed up to high level, are necessary to interpret an experimental data.  

    In spite of impossibility to calculate these coefficients so as a scattering amplitudes and, in turn, cross-sections, it seems that obtaining the set of points or areas of fulfillment conservation laws can not be damaged for any scheme of physical theory. Asymmetry of the equations for $j_1$ and $j_2$ in (10) with respect to a potential variations together with its different initial values, entered the incident beam of particles, may lead to essential different particles trajectories. As a consequence, it means that different particles originating from scattering area will have different directions of motion with respect to direction of incident beam. It allows one to use an ordinary definition of the differential cross-section, but a process of scattering would be subjected to the group-theoretic scheme of quantum mechanics.

\section{Conclusion}
\pst

     Having an aim to make more clear the question in title, we have compared here two approaches to description of quantum phenomena 
by means of example of the simplest problem of quantum theory - the problem on quantum particle transmission above the stationary unidimensional potential step.

     One approach is the generally accepted one, and the problem solution was reduced here in accordance with
\cite{6},
including some negligible extensions in the framework of such solution only.

     Another approach may be named as a consecutive group-theoretic one.
The subject is appeared from discovery of the fact that the contemporary schemes of quantum mechanics are not consecutive group-theoretic ones.
It is evident extremely expressively from the circumstance that there are used two binary operations over a set of the wave functions or spinors transformations.

    The consecutive transformations are the Markovian ones, and there is defined the operation of multiplication over the set of such transformations. Respectively, the groups of  transformations have to be multiplicative ones, they may contain non-commutativity. The last circumstance is especially important for quantum mechanics.

    Evidently that quantum theory can not not be restricted with only consecutive transformations. Independent, or  alternative, processes require to use some kind of composition of alternative propagators. There was introduced the superposition principle (see
\cite{6}
 for more details) in the general accepted schemes of quantum mechanics. The last one may be named more exactly: the Euclidean superposition principle.

    Thus, the second binary operation is appeared over the transformations set - the addition, besides multiplication. It leads immediately to rejection of the group-theoretic tracing of the quantum theory. Moreover, an additive groups do not contain non-commutativity, and one discovers that propagators are existing as if they would be in two different spaces with essential different Gaussian curvatures: one of them has non-zero curvature - for consecutive and non-commutative in general case propagators, and another with zero Gaussian curvature - for alternative, having the same rights, i.e. commutative,  propagators. Possibly,  it may be recognized as the complementary geometrical argument on impossibility of tracing the quantum theory as the consecutive group-theoretic one on the base of combination of Markovian properties of propagators with the ordinary (Euclidean) superposition principle for them.

      However, it was found later that solution of the problem on the composition principle for the non-commutative Markovian propagators on the group is quite insufficient to tracing the quantum mechanics as the consecutive 
group-theoretic physical theory. The problem had been arisen even if 
the non-Euclidean superposition principle for alternative propagators is used.

     The matter consists of following circumstances. The Noether theorems establish one-to-one correspondence between fulfillment of the conservation laws and the group-theoretic requirements to transformations
\cite{7}.
It means the theory has to be traced as the group-theoretic one to fulfillment of all conservation laws, on the one hand. On the other hand, it is {\it Unthinkable} for any theory that all conservation laws may be fulfilled under absence of the complete set of observables. Therefore, it is necessary  to find such complete set beforehand
\cite{14,8,9}.

    Now it is difficult to formulate a sufficient requirements for tracing the fundamental physical theory as the consecutive mathematical group-theoretic scheme. But it seems to be clear now that at least two necessary elements have to be included into such quantum mechanics scheme. They are the non-Euclidean superposition principle and the complete set of observables, or the Hermitian forms. 

    It can be drawn some conclusions with respect to foundation of these elements.

    First of all, the additional two Hermitian forms,
 $\psi ' \psi^{*'}$ and $\psi \psi^{*'} + \psi ' \psi^*$,
 included into the scheme in addition to two well known ones,
$\psi \psi^*$ and $i(\psi \psi^{*'} - \psi ' \psi^*)$,
allow one to hope that they will take a role of the hidden parameters. Hidden parameters were forbidden by von Neumann theorem
\cite{5}. 
However, his proof was considered  to be doubtful, for example in
\cite{2}),
 and also discussed in turn by J. Bell (see
\cite{27,28}). 
The theorem contains some conditions which say that introduction of the hidden parameters is certainly impossible without essential modifications of the existing theory.

    One may note that neither the  conventional quantum mechanics schemes, nor the theorem do not include some kind of completeness condition for observables. In opposite case such condition would be a proof of the theorem, one can not introduce something into the complete set of observables as far as all of them would be there at this time. Moreover, such completeness condition would allow an experimental verification in principle, similar to the Bell inequalities. 

     In the first place, the quad of Hermitian forms described above (see also
\cite{8,9,14,19})
is well known during many years under the name of the Stokes parameters, it may be seen from comparison of the expressions (7) with similar expressions in
\cite{25}.
Besides, the Hermitian forms $j_1, j_2, j_3$ from (6) were used in 
\cite{29}
as three components of spin. Thus, two additional Hermitian forms (or only one!) entered the quad, can not be named as a {\it hidden} parameters in no way, they would be nothing but {\it forgotten} ones.

     In the second place, the additional Hermitian forms are constructed by means of only the same objects, $\psi$ and $\psi'$, already used in the theory to construct the pair of the Hermitian forms, which are also included into quantum mechanics. It is difficult to wait that they may be a cause of {\it essential} modifications of quantum theory. For example, the derivative of the general accepted "probability density" is equal to the unacceptable Hermitian form $j_2$ (see the text below of the expression (39)). 

   During many years quantum mechanics is accompanied by the probability concept. During the same years the problem of hidden parameters is discussed in the papers devoted to foundations of quantum mechanics. One may note at the same time that during the same period the probability density  $\psi \psi^*$ was not accompanied by the value $\psi ' \psi^{*'}$
as long as the (convective) probability density current 
$i(\psi \psi^{*'} - \psi ' \psi^*)$ was not accompanied by the (diffusion) probability density current $\psi \psi^{*'} + \psi ' \psi^*$.

   It should be also mentioned a general confidence that all conservation laws are fulfilled in quantum theory in just the same way as almost  general confidence that the Noether theorems are also contained there. Of course, 
these theorems may be considered as the mathematical tool for realization 
of conservation laws in theory, but it is necessary to fulfill some conditions.
That is to say the quantum theory has to be traced as the consecutive mathematical group theory. In opposite case the first circumstance is not 
supported by the second one, as it takes place in quantum theory now.

    It is probably at the same time, that inclusion of the complete set of the Hermitian forms into its scheme may lead to displacement of the probability from the scheme of quantum mechanics. 

    The authors
\cite{2} 
had put and discussed the question: 
 can quantum-mechanical description of physical reality be considered complete? All schemes of quantum mechanics did not contain earlier and 
do not contain now even some kind completeness condition, similar to (8).
Only two of four Hermitian forms were included into consideration, it is obviously insufficient. This fact may explain necessity of use a probability concept in these schemes. Inclusion of complete set of observables consisting of four Hermitian forms allows one to hope at least for their coexistence with this concept.

   To make the question in title more clear it was considered the simplest problem of quantum mechanics here. If incompleteness is the general property of the theory then it has to be appeared also in the simplest case. For example, the authors
\cite{6} 
solving the problem on the potential step had used the term "current of the probability density".

    Here it had been investigated such case, and now it can be said that the probability concept is excessive one for quantum theory
 \footnote{For example, classical electrodynamics is also non-group-theoretic theory, but, containing the complete set of observables - Stokes parameters, does not need a probability concept at all.}.
 Inclusion of the complete set of the Hermitian forms into the scheme leads to complete description of the object described by the Schroedinger equation, under known initial conditions, of course.

     The concept of the "probability" had been developed up to extremely high level in quantum mechanics, especially as philosophical one. Maybe, it would be turned out to be useful, and even necessary. But it seems to be clear now that it is necessary for it to prove, similar to the von Neumann theorem, that use of the additional Hermitian forms, constructed on the base of the same values as already used, is forbidden due to some extremely important reasons.

   Sometimes it seems that to understand quantum mechanics means to understand why it is possible to use $\psi \psi^*$ and $i(\psi \psi^{*'} - \psi ' \psi^*)$ but it is impossible to use $\psi ' \psi^{*'}$ and $\psi \psi^{*'} + \psi ' \psi^*$ at the same time.

    The non-group-theoretic schemes of quantum mechanics, absence of the complete set of observables on a general background of insignificant role of the Noether theorems in contemporary schemes of quantum mechanics leading to probability and hidden parameters concepts there, accompanying by violation of some conservation laws, force to express a suspicion formulated in short as "probability probably hides perpetuum mobile".

     It is impossible to bypass the double-slit experiment playing a great role in quantum mechanics. Estimating its significance the author
\cite{1}
even had said that this experiment contains all enigmas of quantum mechanics up to 100 percents. One would like to agree with this statement, both from experimental point of view and also from theoretical one, but taking into account two additions: the slits have to be point-like ones in the experiment and they have to be placed onto the boundary of two media in the theory. The first circumstance excludes unnecessary paths from measurements, and the second one includes non-commutativity into its theoretical description.

    In accordance with mentioned above, one may say that inclusion of additional parameters into the scheme of quantum mechanics is a subject of exigent necessity. This circumstance is equivalent to fulfillment of all conservation laws.

   What is a cause that quantum mechanics was found to be constructed as the non-group-theoretic theory? Taking into account that the necessary similar complete set of the Hermitian forms was well known under the name of the Stokes parameters in just the same way as the group theory and the Noether theorems long before of quantum mechanics creation, it should be considered to be enigmatic. The reasons of this situation are unclear, the more so as such significant for any principal physical theory the Noether theorems are not mentioned in such significant for quantum mechanics books as
\cite{5,6}.

      "Mathematical Foundations of Quantum Mechanics" [5] had been adopted as the foundations of the contemporary schemes of quantum mechanics. This is  based, in turn, on the set, which forms the Hilbert space, having its own axiomatics. The Noether theorems, connecting symmetries of space and transformation properties of physical variables with conservation laws, require the set, which forms groups, also within its own axiomatics.

    Therefore it is relevant to put a number of questions. 

   If the quantum theory is traced in the Hilbert space and conservation laws need a set named groups, then the very modest requirements consist of some compatibility of these two sets. Are these sets consistent now? Does the mathematical statement which may be considered to be equivalent to the Noether theorems exist in the Hilbert space? What is the Gaussian curvature of the Hilbert space and does this curvature may have a constant negative value? The last one seems to be quite necessary due to noncommutativity of transformations in quantum mechanics.

   Does the von Neumann theorem on hidden parameters contain an exclusion principle for inclusion
$\psi ' \psi^{*'}$ and $\psi \psi^{*'} + \psi ' \psi^*$ into quantum theory?
 Is the same exclusion principle included into any other books on quantum mechanics, especially devoted to its foundations? If yes, then why? If not, then why they are not included?

    Finally, let us mention shortly two questions. The first one is on opportunity of coexistence a probability interpretation in quantum mechanics under presence of complete set of observables - four Hermitian forms. The second one is a question on opportunity of experimental demonstration of this set existence.

     The fact of formulation and proving by von Neumann  of the  theorem     
on impossibility to introduce hidden parameters into quantum theory says us that its author supposed their coexistence to be impossible. It is unclear at the same time what constructive elements this theorem had inserted into quantum theory
\cite{27,28}.

    However, it is extremely doubtful that some kind of interpretation, even probabilistic one, may be sufficient foundation for conservation laws breakdown, fulfillment of which is unthinkable under absence of the complete observables set. An existence of last one does not depend on presence or absence of some interpretation, therefore coexistence is possible.

    Relating to experimental demonstration of the set existence one may say that the Hermitian forms $j_1, j_2, j_3$, constructed on the base of spinor components, similar to expression (6), are used in quantum mechanics long ago 
\cite{29}
 as the spin components. Although "spin is  key and still not fully understood  property of matter" 
\cite{30}, 
it had entered quantum mechanics long ago and stably. Therefore one may hope that spin appearance in experimental investigations may be  found to be also a partial experimental foundation for the complete set of observables existence, three of four Hermitian forms of the complete set of which make up in turn this spin concept.

\end{document}